\documentclass[5p,12pt]{elsarticle}

\usepackage{siunitx}
\usepackage{float}
\usepackage{amsmath}
\usepackage{tikz}
\usepackage{tikzscale}
\usetikzlibrary{calc}
\usepackage[hyphens]{url}

\usepackage{amssymb}
\usepackage{relsize}    
\usepackage{xspace}     
\usepackage{todonotes}  

\journal{Astroparticle Physics}

\begin{document}

\begin{frontmatter}



\title{Probing the cosmic ray energy spectrum at $10^{12}$--$10^{16}$ eV with two HiSPARC scintillators}


\author[nikhef]{K. van Dam\corref{cor}}
\ead{kaspervd@nikhef.nl}
\author[nikhef,twente]{B. van Eijk}
\author[nikhef]{J.J.M. Steijger}

\address[nikhef]{Nikhef National Institute for Subatomic Physics, Science Park 105, 1098 XG Amsterdam, The Netherlands}
\address[twente]{Faculty of Science and Technology, University of Twente, 7500 AE Enschede, The Netherlands}

\cortext[cor]{Contact information}

\begin{abstract}
The high school project on astrophysics research with cosmics (HiSPARC) employs a large number of small detection stations that sample the footprint of extensive cosmic ray air showers. The majority of these stations has two 0.5 \si{\meter\squared} scintillation detectors. A new method is presented which enables probing the cosmic ray flux with a single two-scintillator station in five energy decades at $10^{12}$, $10^{13}$, $10^{14}$, $10^{15}$ and $10^{16}$ eV. The method is based on the energy dependence of the distribution of the number of particles passing through a single detector. A relatively short data taking period of approximately one month is sufficient to probe this energy range. The flux values agree well with measurements by other experiments. For the first time, the cosmic ray flux at $10^{12}$ and $10^{13}$ eV is derived at sea level.
\end{abstract}

\begin{keyword}
HiSPARC \sep Cosmic rays \sep Scintillation detector \sep Cosmic ray energy spectrum


\end{keyword}

\end{frontmatter}


\section{Introduction}
Cosmic rays are charged particles from space with energies ranging from GeV to hundreds of EeV. The majority of cosmic rays are protons (90\%) and helium nuclei (9\%) with a remainder of heavier elements \cite{gaisser2016}. The cosmic ray flux drops rapidly with increasing energy. The flux as function of energy ($E$) in the range from GeV to $10^{14}$ eV is approximately given by \cite{physicalreviewd2012}:
\begin{equation}
\label{eq:crflux}
I(E) \approx 1.8 \cdot 10^4 \ (E/1\ \text{GeV})^{-\alpha} \frac{\text{\# of nucleons}}{\text{m}^2\ \text{s sr GeV}}
\end{equation}
with $\alpha \approx 2.7$. At an energy of $\sim$$10^{15.5}$ eV the spectrum steepens. This region is known as 'the knee'. At $\sim$$10^{18.5}$ eV the slope of the spectrum becomes less steep, this is called 'the ankle'. Cosmic rays with energies below $\sim$$10^{14}$ eV can be measured using high altitude balloons or spacecraft. For higher energies this approach becomes unfeasible because of the low cosmic ray flux, the limited collection area and lifetime of space based instruments. Above $\sim$$10^{14}$ eV cosmic rays are studied using the Earth's atmosphere. When a cosmic ray enters the atmosphere it will most likely interact with a nitrogen or oxygen nucleus. In this interaction new particles may be produced which, in turn, will interact with other atmospheric particles. The particle generation continues until there is insufficient energy left for the creation of additional particles. This phenomenon is known as the development of an extensive air shower (EAS). The particle footprint of an EAS at the ground is sampled using detector arrays. The direction and energy of the cosmic ray can be reconstructed from particle arrival times and multiplicities. Also atmospheric radiation generated by the EAS, such as atmospheric Cherenkov and fluorescence, and radio waves, can be used to reconstruct the shower size and direction \cite{gaisser2016}. Cosmic rays with energies below $\sim$$10^{14}$ eV do not result in large air shower footprints. The particle densities are too small to reconstruct the direction and energy of the primary cosmic ray.

\begin{figure*}
	\centering
	\includegraphics[width=140 mm]{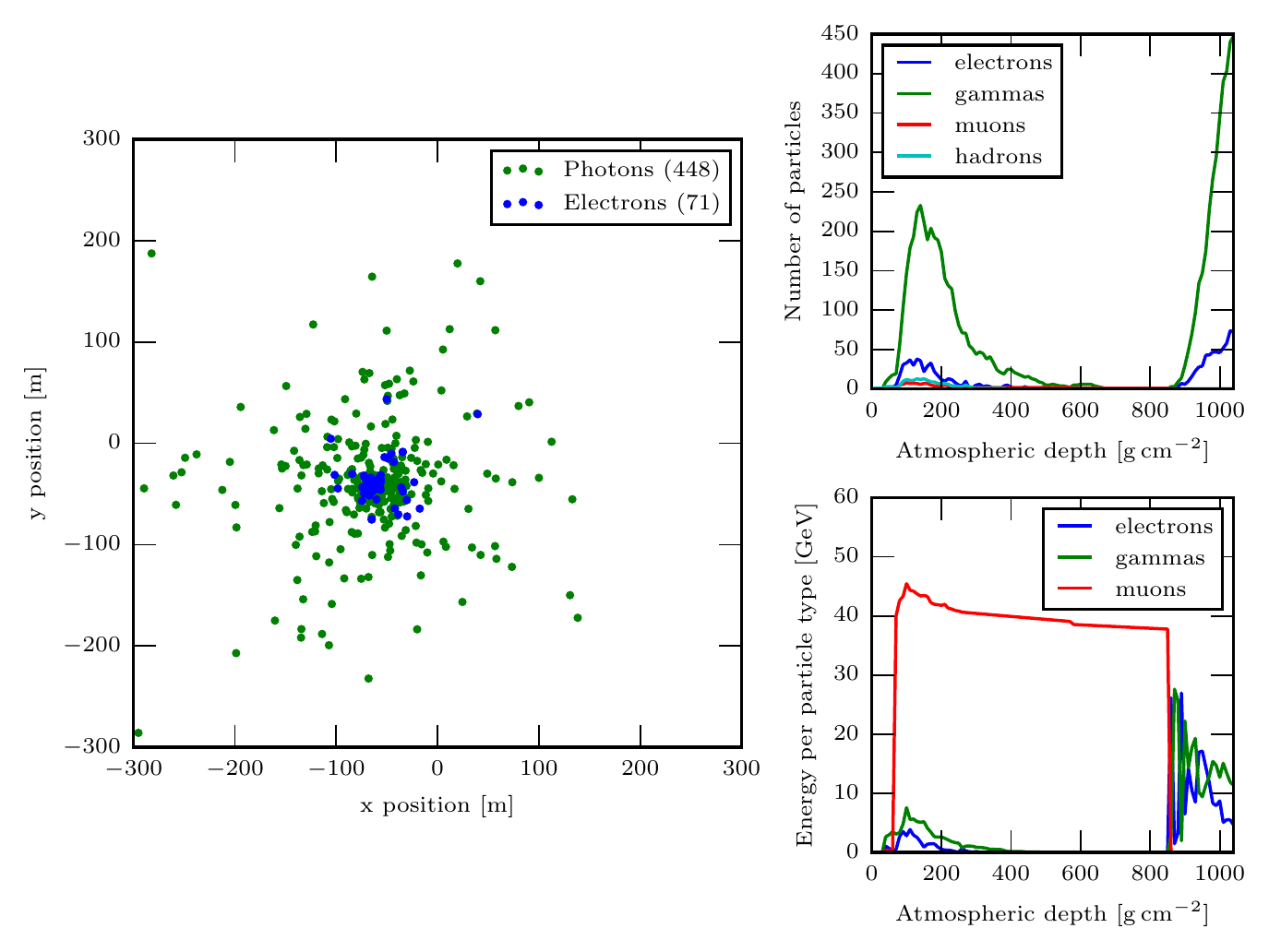}
	\caption{Example of a muon decay induced shower in a $10^{11}$ eV proton CORSIKA \cite{corsika1998} simulation. The upper right figure shows the longitudinal profile, i.e. the number of particles per particle type as function of atmospheric depth. The original shower dies out and, at a depth $\sim$900 \si{\gram\per\cm\squared}, a second (very large) shower emerges. The bottom right plot shows the energy per particle type as function of the atmospheric depth. In one of the first interactions, $\sim$40 GeV of the available 100 GeV is transferred to a single muon. At a height of 1.6 km (855 \si{\gram\per\cm\squared}) this muon decays and a large part of its energy is passed on to an electron (positron) which ignites an e.m. shower. The left figure shows the e.m. shower footprint at the ground. The position of the shower core is slightly shifted from the direction of the primary proton as particles are absorbed and neutrino(s) remain undetected.}
	\label{fig:mini_shower}
\end{figure*}

If an even lower energy cosmic ray (e.g. $10^{11}$ eV) hits the Earth's atmosphere, the air shower dies out before reaching the ground. The majority of shower particles are absorbed at high altitude. Only energetic remnants, of which mainly muons (and neutrinos), are able to reach ground level. The flux of perpendicularly incident muons at sea level is $\sim$70 \si{\per\meter\squared\per\second\per\steradian} above 1 GeV \cite{physicalreviewd2012}. If an energetic muon decays before reaching the ground, the decay electron (positron) emits gamma radiation due to Brems\-strahlung. Energetic gamma rays will interact with matter via pair production. The two processes (Brems\-strahlung and pair production) generate an electromagnetic shower. Figure \ref{fig:mini_shower} shows an example of such a shower in a $10^{11}$ eV proton CORSIKA \cite{corsika1998} simulation. The upper right figure shows the longitudinal profile, i.e. the number of particles per type as function of atmospheric depth. The original shower dies out and, at a depth $\sim$900 \si{\gram\per\cm\squared}, a second, electromagnetic (e.m.), shower appears. The bottom right plot shows the energy per particle type as function of the atmospheric depth. In one of the first interactions, $\sim$40 GeV of the 100 GeV proton is forwarded to a single muon. At a height of 1.6 km (855 \si{\gram\per\cm\squared}) this muon decays and a large fraction of its energy is transferred to the electron (positron) which initiates an e.m. shower. The left figure shows the footprint created by the shower at the ground. The particle density is very small. The position of the shower core is shifted with respect to the original direction of the cosmic ray proton due to absorption of the shower particles other than the muon, and neutrino(s) escaping detection. Muon decay is the dominant source for electrons at sea level. The total flux of perpendicular incident electrons and positrons above 10 MeV is $\sim$30 \si{\per\meter\squared\per\second\per\steradian} \cite{physicalreviewd2012}.

The high school project on astrophysics research with cosmics (HiSPARC \cite{vandam2020}) employs a large number of small EAS detection stations. The stations are predominantly hosted by high schools. Each station consists of two or four scintillators connected to an electronics unit that digitizes the analogue PMT signals. A station uses a GPS receiver for accurate timing and position information. The majority of high schools employs a two-detector station. As neighboring schools often also have HiSPARC equipment, multiple stations are used to sample the same EAS. Data sets are combined to reconstruct the air shower.

A new method is presented to enable the determination of the cosmic ray flux at energies below $10^{14}$ eV using two scintillators at sea level. The method does not require the full reconstruction of individual EASs but is based on measuring the energy dependence of the distribution of the number of particles passing through a detector.

First, the experimental setup is introduced after which the new method is detailed. Finally, the results are compared with measurements by other experiments.

\section{HiSPARC station}
\begin{figure*}
	\centering
	\includegraphics[width=140 mm]{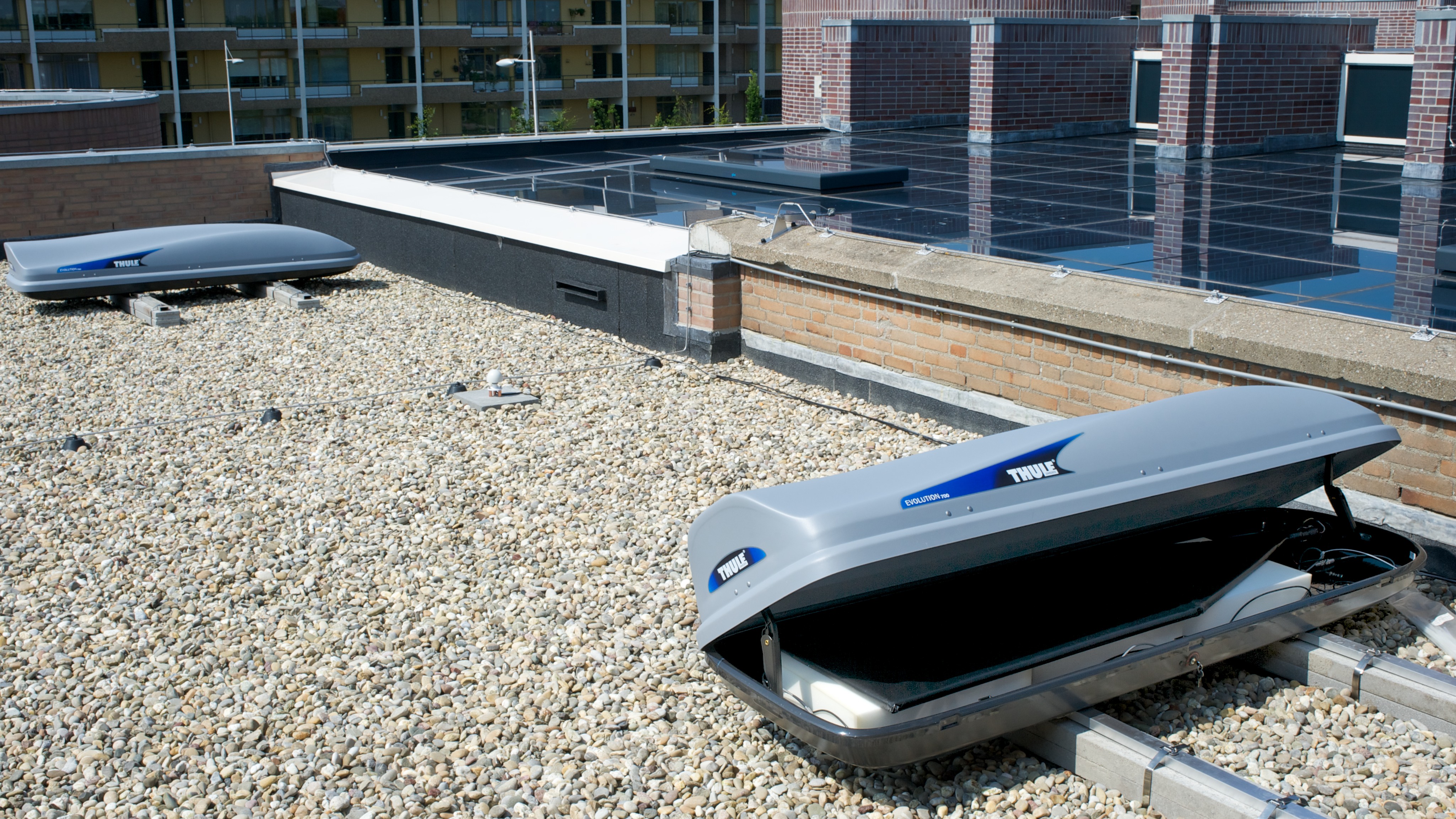}
	\caption{A two-detector HiSPARC station with two scintillators placed in roof boxes mounted on top of a roof. The detector separation is typically $\sim$5$-$10 m. In between the two detectors a GPS antenna is placed for accurate timing and location information. Photo courtesy of Arne de Laat.}
	\label{fig:twodetectors}
\end{figure*}

A HiSPARC detector consists of a rectangular scintillator (100 cm $\times$ 50 cm $\times$ 2 cm) glued to a light-guide which is attached to a PMT. The detector is made light tight with a thin aluminum foil and pond liner. The assembly is placed inside a roof box and mounted on top of a roof (figure \ref{fig:twodetectors}). When a charged particle traverses the scintillator, it creates a light pulse which is converted into an electric pulse by the PMT. This pulse is sampled and digitized at 400 MHz. The PMT is calibrated such that single charged particles generate pulses with an amplitude of $\sim$150 mV. A 30 mV threshold rejects PMT noise and the response to low energy gamma rays. The detector response is simulated using GEANT4 \cite{GEANT42016} and verified against experimental data. The detection efficiency for minimum ionizing particles (MIP) is $\sim$100\%. The detector has a much lower sensitivity for gamma rays in EASs ($\sim$10\%) \cite{vandam2020}.

To discriminate against single particle sources, HiSPARC applies the 'coincidence method'. By using two (or four) detectors and selecting coincident PMT pulses (i.e. with a time difference smaller than 1.5 \si{\micro\second}) the majority of random coincidences are rejected. If two or more PMT pulses exceed the threshold within the trigger window, the pulses are stored. Random coincidences occur due to gamma rays from decays of nearby environmental radionuclides \cite{vandam2020b} and muons generated by low energy cosmic rays of which the other components of the air shower are absorbed in the atmosphere, etc. A rare event like the one in figure \ref{fig:mini_shower} has a chance to result in a trigger.

The distance between the detectors in a two-detector station varies per setup. In this paper a detector separation of 4.95 meter is used. In a four-detector station two electronics units are combined in a master-slave configuration. Similar trigger conditions apply as for a two-detector station. In a four-detector station the direction can be triangulated with a resolution of $\sim$6\si{\degree}. An estimate of the shower energy can also be made. An extensive description is presented in \cite{vandam2020}.

\section{Pulse integral distribution (PID)}
The value of the PMT pulse integral, i.e. the area under the pulse, is directly proportional to the number of scintillation photons that reach the PMT, and thus to the number of particles traversing the detector (MIPs). The energy loss of a MIP follows Landau's theory \cite{landau1944}. The Landau distribution has a peak at the most probable energy loss with a pronounced tail towards higher energies. If multiple particles simultaneously traverse a detector, the energy loss is described by a sum of Landau distributions. The shape of the pulse integral distribution is then determined by the number of MIPs that traverse the detector within the trigger time window ($1.5$ \si{\micro\second}). The energy of the primary cosmic ray is directly proportional to the number of EAS particles in the footprint at the ground.

\subsection{Analyzing the PID}
\begin{figure}
	\centering
	\includegraphics[width=90 mm]{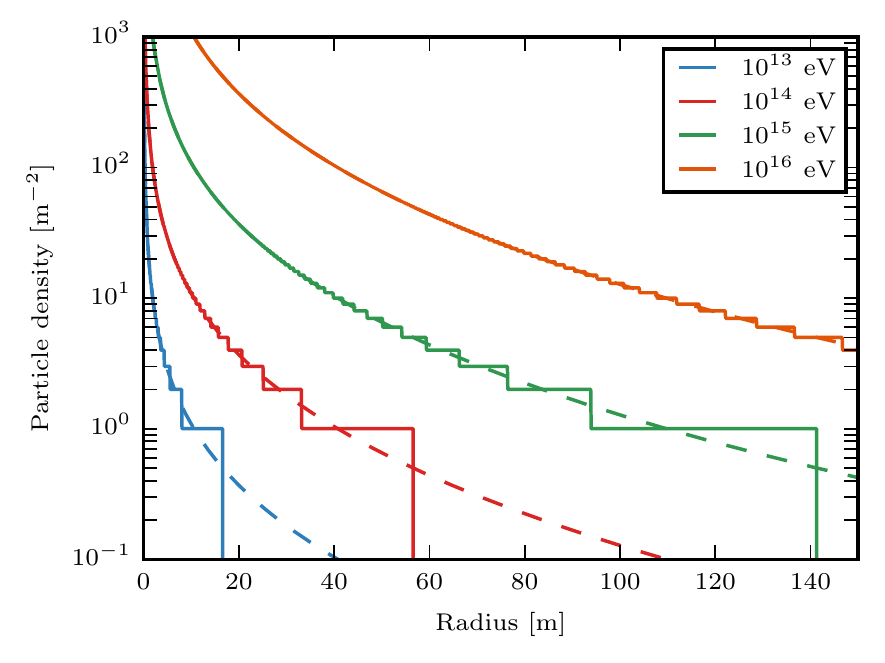}
	\caption{The dashed lines show the lateral density profiles for $10^{13}$ (blue), $10^{14}$ (red), $10^{15}$ (green) and $10^{16}$ eV (orange) showers described by eq. \ref{eq:agasa}. The solid lines show the multiplicity rounded to the nearest integer. In the calculation only EASs from the zenith are considered.}
	\label{fig:discrete_NKG}
\end{figure}

\begin{figure}
	\centering
	\includegraphics[width=90 mm]{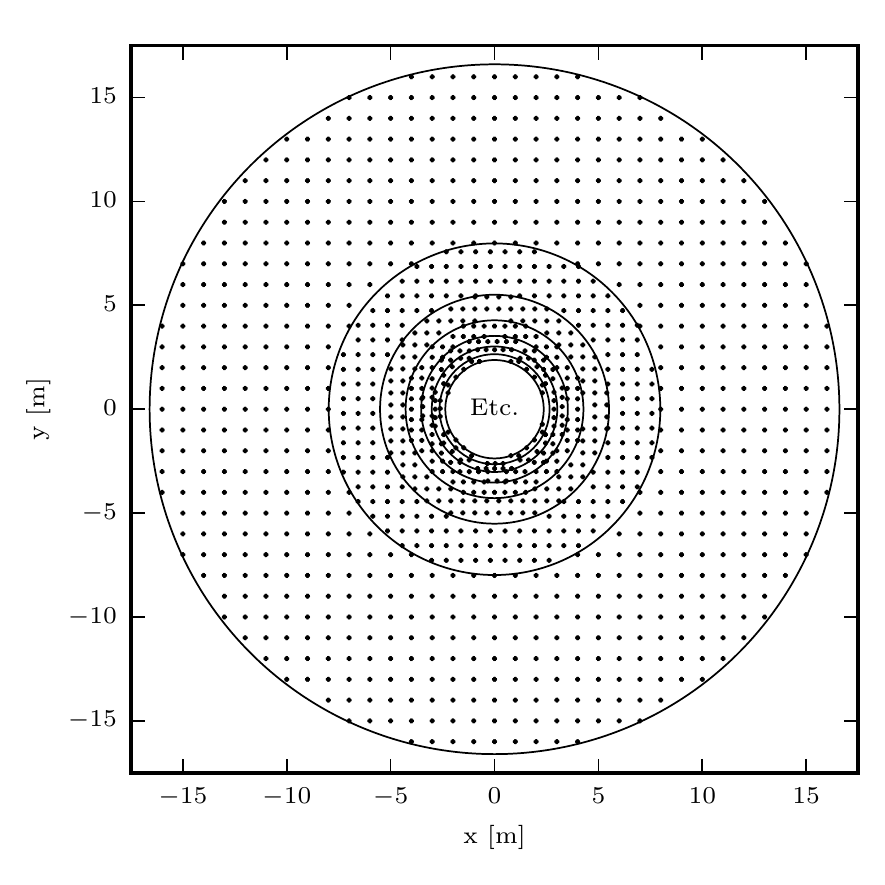}
	\caption{Example of a $10^{13}$ eV perpendicular incident EAS footprint. Each dot represents a single particle. Within each ring the particle density is taken constant. When a detector with a size of $1 \times 1$ \si{\meter\squared} is randomly projected on the density map, the probability to detect $x$ number of particles scales with the surface area of each ring. For $x=1$ the probability is equal to the surface area of the outermost ring divided by the surface area of the entire disc.}
	\label{fig:circles}
\end{figure}

In a first approach all cosmic rays are assumed to stem from the zenith. The particle densities can be approximated by the lateral density profile from AGASA \cite{takeda2003}:
\begin{align}
\label{eq:agasa}
S(r) \propto &\left( \frac{r}{r_\textup{M}} \right)^{-1.2} \left( 1+\frac{r}{r_\textup{M}} \right)^{-(\eta-1.2)} \nonumber \\
&\times \left[ 1 + \left( \frac{r}{1000} \right)^2 \right]^{-0.6}
\end{align}
with distance to the shower core $r$ in meter, $\eta=3.84$ and Moli\`ere radius $r_\textup{M}=91.6$ m. The energy associated with this lateral density profile is given by \cite{dai1988}:
\begin{equation}
\textup{E} = a \cdot 10^{17} \cdot (S(600))^{b} \quad [\textup{eV}]
\end{equation}
with $a=2.03$ and $b=1.0$. The dashed lines in figure \ref{fig:discrete_NKG} show EAS particle densities as function of distance to the shower core for primary energies of $10^{13}$ (blue), $10^{14}$ (red), $10^{15}$ (green) and $10^{16}$ eV (orange). Obviously, as the EAS energy increases, the size of the footprint increases as well. Next, the particle density per unit area is rounded to an integer. This is illustrated by the solid lines in figure \ref{fig:discrete_NKG}. As an example a $10^{13}$ eV EAS footprint with a regular pattern is shown in figure \ref{fig:circles}. Within each ring the particle density is constant. Each dot represents a single particle.

\begin{figure}
	\centering
	\includegraphics[width=90 mm]{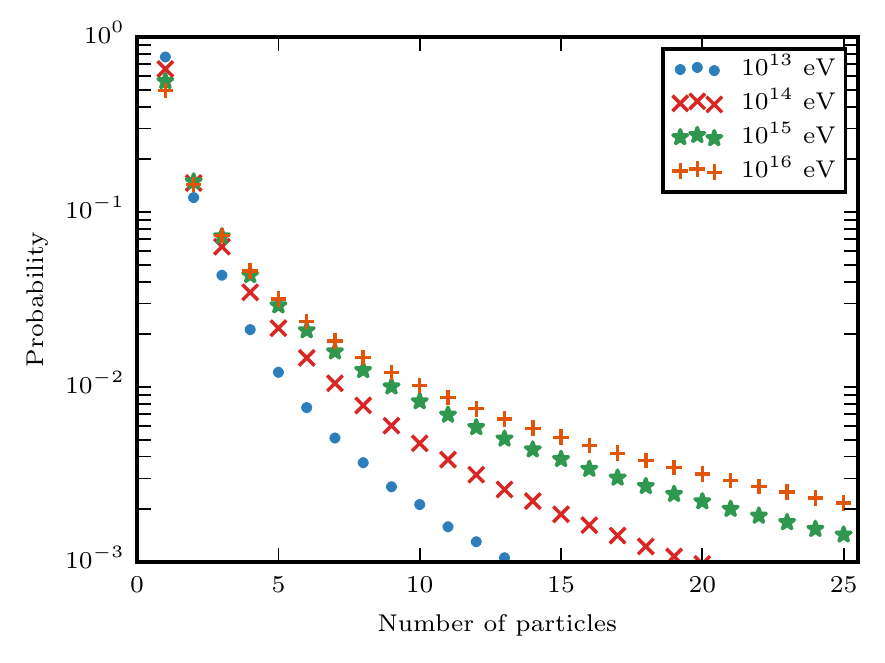}
	\caption{The probability to detect a number of particles in the footprint at a fixed energy. For low energy EASs the probability to detect a single MIP is relatively large compared to higher multiplicities. With increasing energy the slope of the distributions changes significantly.}
	\label{fig:probabilities}
\end{figure}

If a detector with a size of 1 \si{\meter\squared} is randomly projected at the density map, the probability to detect $x$ number of particles scales with the surface area of each ring. For $x=1$ the probability is equal to the surface area of the outermost ring divided by the surface area of the entire disc. This is illustrated in figure \ref{fig:probabilities} which displays the probability density to find a number of particles as a function of EAS energy. Each probability density represents a single energy PID. When many particles simultaneously traverse the detector, a large pulse integral is generated. If only one particle traverses the scintillator, the value of the pulse integral follows a single Landau distribution.

In order to predict the PIDs more accurately, a Monte Carlo simulation is constructed that includes multiple EAS zenith angles and accurate EAS evolution and detector description.

\subsection{Detector separation}
\label{sec:detsep}
The small particle densities at low energy EASs ($\lesssim 10^{13}$ eV) generate a steeply falling PID, i.e. there is a relatively high probability to detect just 1 or 2 particles; higher multiplicities at this energy become rare. Increasing the distance between the detectors then translates into an energy cut. Since the size and shape of the PID directly depends on the number of particles that 'simultaneously' traverse the detector, it provides an indirect measure for the energy of the primary cosmic ray. Figure \ref{fig:discrete_NKG} shows that for $10^{13}$ eV showers (blue curve) the probability to detect particles beyond 20 m from the shower core becomes extremely small. Thus, if two HiSPARC detectors are placed 40 m apart, the EAS detection probability at this energy vanishes. Since it is impossible to select single energy EASs, a PID obtained by a station can be thought of as a sum of single energy PIDs. Introducing an energy cut by increasing the detector separation will affect the shape of the station's PID as single low energy PIDs are excluded from the collection.

\begin{figure}
	\centering
	\includegraphics[width=90 mm]{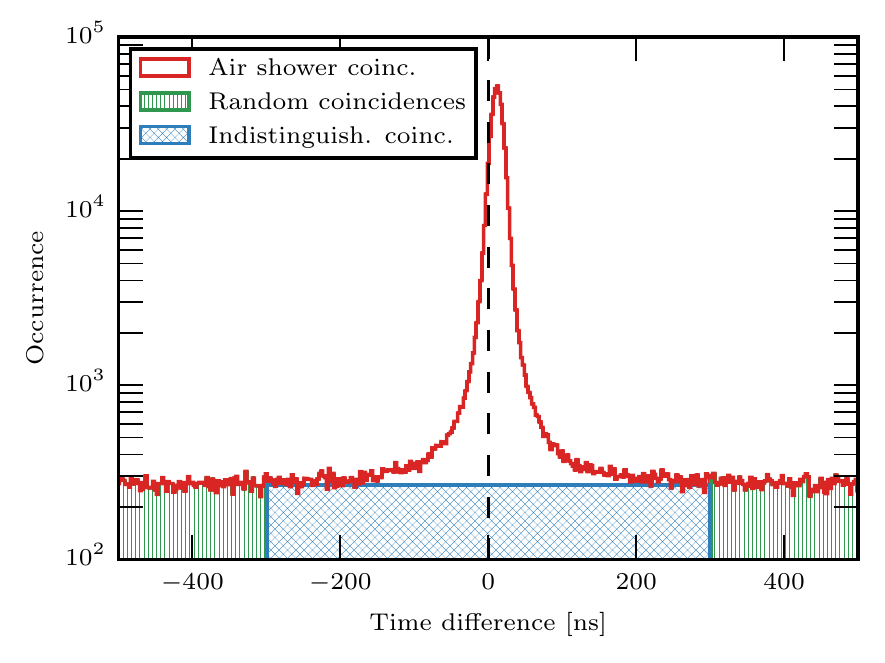}
	\caption{Distribution of arrival time differences between particles that satisfy the trigger condition in a two-detector HiSPARC station (red). The deviation of the peak value from zero indicates a small timing offset between the two detectors. The plateau of events (green striped region) is extrapolated towards small time differences (blue horizontal line) and is the result of random coincidences (PMT noise, etc.). For time differences smaller than 300 ns (blue crossed region) the random coincidences are indistinguishable from air shower events. Their contribution is however small.}
	\label{fig:timediff}
\end{figure}

Figure \ref{fig:timediff} shows the measured distribution of arrival time differences for particles in the two detectors. The times at which the PMT pulses exceed the threshold are taken as arrival times. The plateau of random coincidences (green striped region) extends to small time differences (blue horizontal line). The peak is caused by particles belonging to EASs. For time differences smaller than 300 ns the random coincidences are obviously indistinguishable from EASs. However, the PID distribution obtained by selecting random coincidences can be subtracted (proportional to the number of background events - blue crossed region) from the PID for time differences smaller than 300 ns.

\begin{figure}
	\centering
	\includegraphics[width=90 mm]{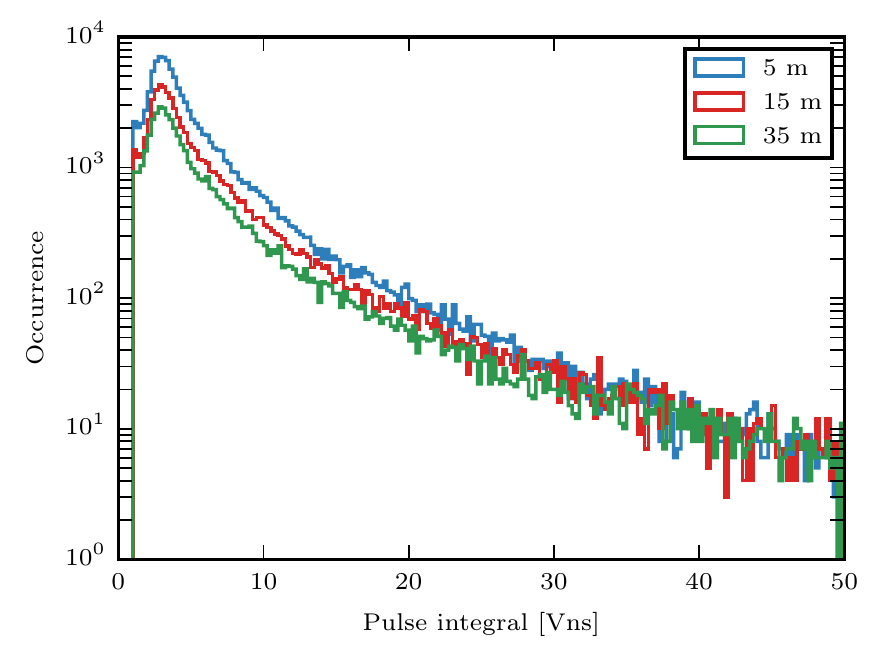}
	\caption{PID for different detector separations. By increasing the distance between the detectors, the station will not trigger on low energy EAS footprints due to their small particle density (Fig. \ref{fig:probabilities}). The number of single MIP detections decreases proportionally to the increasing detector separation. The number of large pulse integrals (multiple simultaneous particles) traversing the detector stays approximately constant since these events are caused by high energy showers which are still detected at large detector separations.}
	\label{fig:pi_distances}
\end{figure}

A small experiment has been carried out in which the PID is measured for different detector separations (5, 15 and 35 m). 
The result is shown in figure \ref{fig:pi_distances}. The number of single MIP events decreases with increasing detector separation. The number of large pulse integrals stays approximately constant since these events are caused by high energy showers.

\subsection{Systematic uncertainties}
\begin{figure}
	\centering
	\includegraphics[width=90 mm]{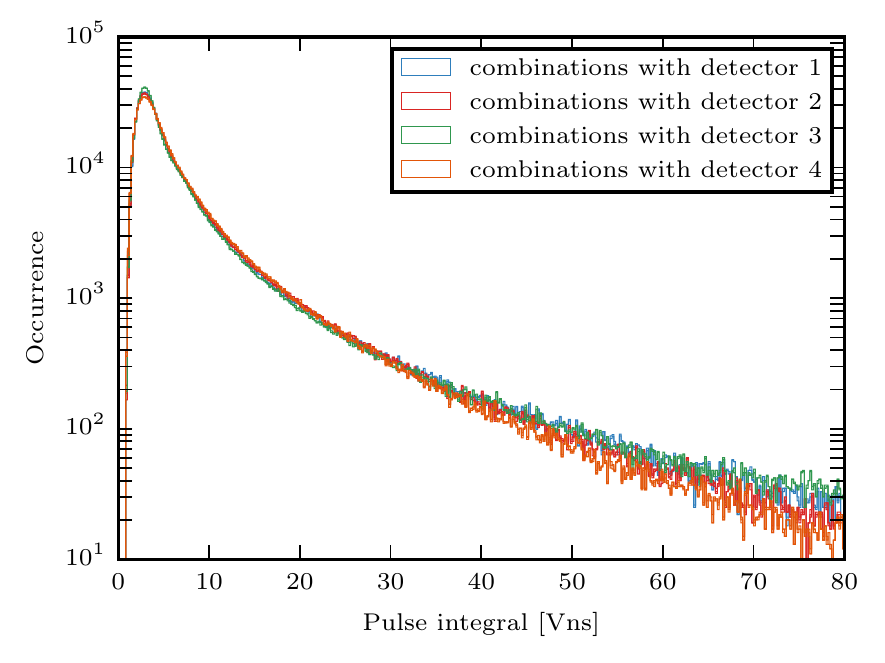}
	\caption{PIDs of ten detector combinations in a four-detector (diamond shaped configuration) station. Each color represents one detector. Per detector two or three PIDs are displayed. The detector separation for each combination is 10 m. Combinations with the same detector (e.g. red lines) are very similar. The difference between the detectors is larger, especially at large pulse integrals (see orange and green lines).}
	\label{fig:systematics}
\end{figure}

The relation between the shape of the PID and the energy of the primary cosmic ray can then be used to probe the cosmic ray flux at a certain energy range. To quantify systematic differences between detectors, the PIDs of detector combinations in a four-detector station (diamond shaped configuration) \cite{vandam2020} have been compared. Each detector in this station can be paired with three other detectors. The four sides of the diamond are 10 m. The long diagonal combination is discarded. This results in ten PID distributions. Figure \ref{fig:systematics} shows the ten distributions of the four detectors. All combinations with 'detector one' are displayed in blue, all combinations with detector two are displayed in red, etc. The PIDs obtained with the same detector are very similar. The difference between detectors is larger and becomes more visible at large pulse integrals (see orange and green lines). This indicates that there are instrumental differences caused by variations in the number of scintillation photons reaching the PMT, large pulse PMT response, etc. After closer inspection, systematic differences between the PIDs up to $\sim$35 Vns ($\sim$12 MIPs) are small.

\section{EAS Simulations}
A large sample of EASs for primary cosmic rays with an energy of $10^{12}$, $10^{13}$, $10^{14}$, $10^{15}$ and $10^{16}$ eV is generated with CORSIKA. Only proton initiated showers are considered. The showers have been generated with zenith angles ranging from 0\si{\degree} to 60\si{\degree} in steps of 3.75\si{\degree}. For high energy hadronic interactions QGSJET-II \cite{qgsjet2006} was selected. Hadronic interactions below 80 GeV are simulated using GHEISHA \cite{gheisha1985} and electromagnetic interactions are incorporated with EGS4 \cite{egs41985}. The full particle shower was simulated (no 'thinning' \cite{hillas1981}). The location of the EAS shower cores were randomly chosen within a circle with a radius of 150 m. The two-detector station is at the center of the circle. The arrival direction (azimuth and zenith) was chosen isotropically (random points on the surface of a unit sphere). The response of the scintillator and light-guide to particles traversing the detector was simulated with GEANT4 \cite{GEANT42016}. A parameterized PMT response was used \cite{vandam2020}. Finally, the HiSPARC trigger conditions were applied.

\begin{figure}
	\centering
	\includegraphics[width=90 mm]{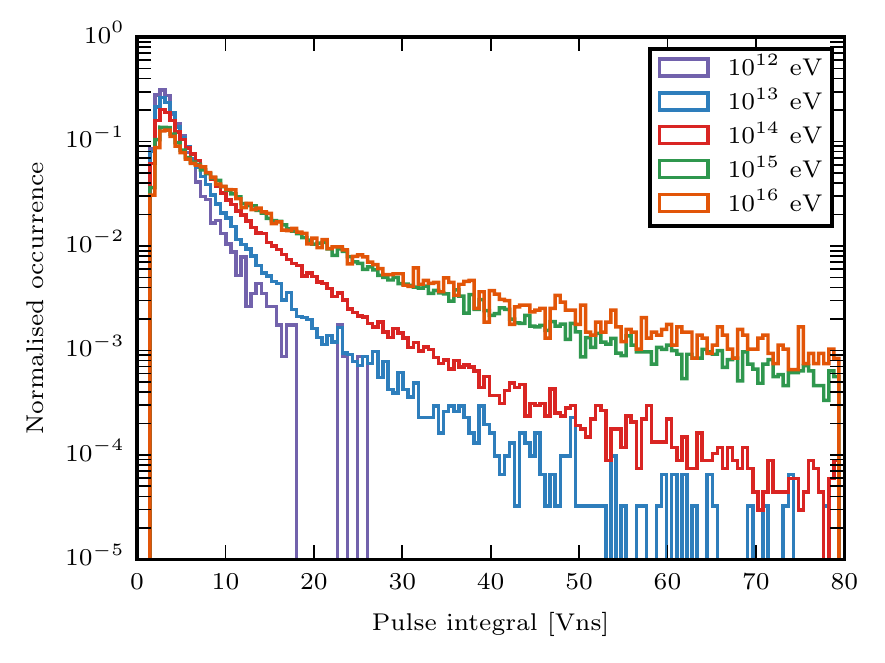}
	\caption{Simulated PIDs at five different energies. As expected from figure \ref{fig:probabilities} the PID depends on the energy of the EASs. Contrary to the model, a full detector simulation is carried out with isotropically selected EASs and uniformly chosen core distances. Only proton induced EASs are considered. The difference becomes evident at large pulse integrals.}
	\label{fig:ph_per_energy}
\end{figure}

Figure \ref{fig:ph_per_energy} shows the normalized, simulated PIDs for the five different energies. At higher energies the relative abundance of large pulses increases considerably. By fitting a linear combination of the simulated distributions to the experimentally observed spectrum, the cosmic ray flux at a fixed energy interval can be estimated.

\subsection{Effective surface area and solid angle}
\begin{figure*}
	\centering
	\includegraphics[width=140 mm]{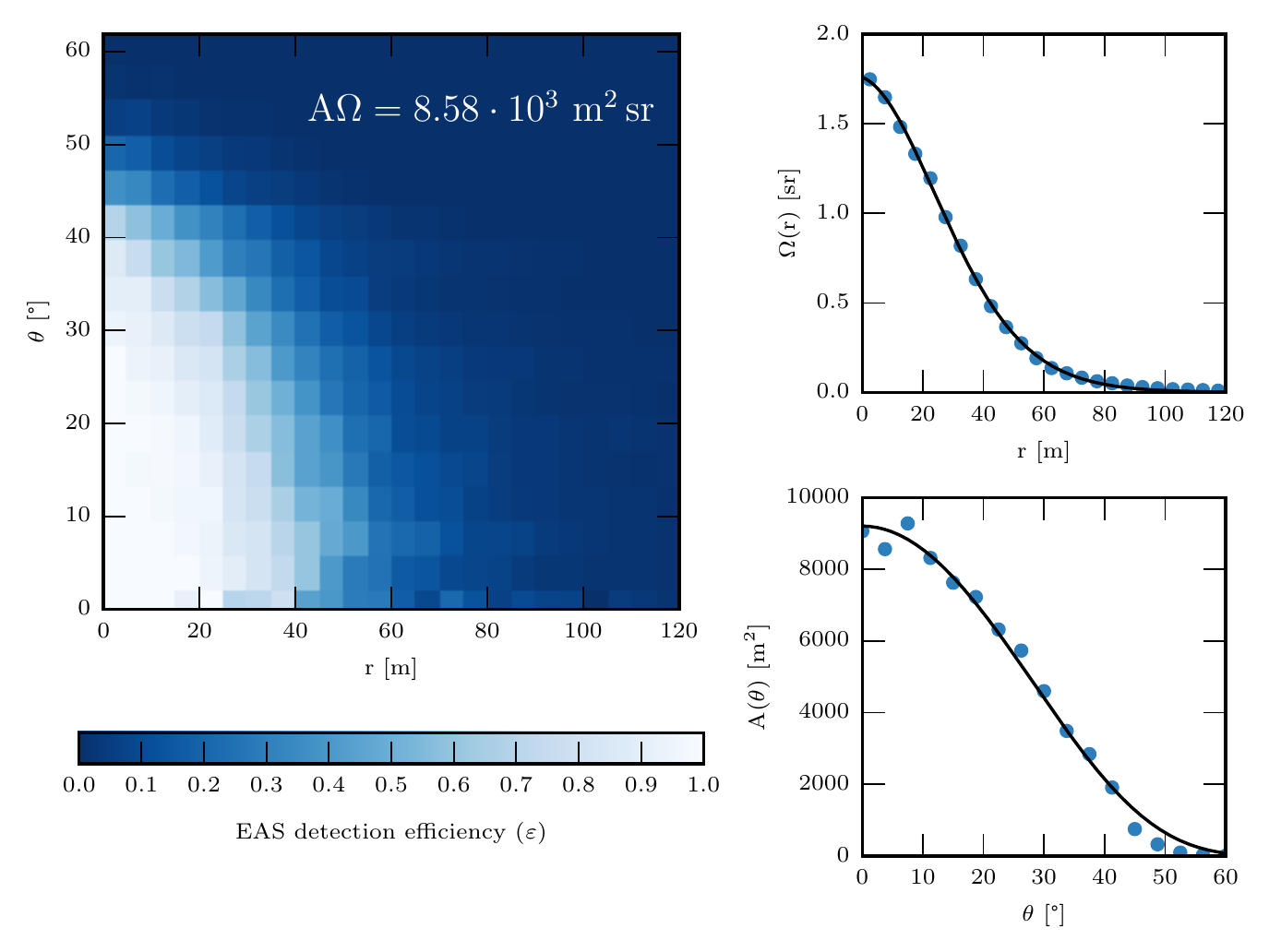}
	\caption{Left: The detection efficiency ($\varepsilon$) for $10^{15}$ eV proton induced showers as function of core distance ($r$) and zenith angle ($\theta$). Top right: the detection efficiency integrated over the zenith angle yields the effective solid angle as function of core distance ($\Omega(r)$, blue points) which can accurately be described using the parametrisation in eq. \ref{eq:omega} (black line). Bottom right: the detection efficiency integrated over the core distance yields the effective surface area as function of zenith angle ($\textup{A}(\theta)$, blue points) can be described by the parametrisation in eq. \ref{eq:a} (black line).}
	\label{fig:solid_angle_surface_area}
\end{figure*}

EASs at large zenith angles result in lower particle densities at ground level because of increased absorption due to the larger path length through the atmosphere. Moreover, the particle density observed in the detection station strongly depends on the distance to the shower core. These two effects need to be quantified in order to calculate the cosmic ray flux. For this an effective surface area and effective solid angle are introduced. The fraction of events that results in a trigger is defined as the 'EAS detection efficiency' ($\varepsilon$). The left plot in figure \ref{fig:solid_angle_surface_area} shows $\varepsilon$ for $10^{15}$ eV air showers as function of the zenith angle and distance to the shower core. The maximum efficiency occurs at small $r$ and small zenith angle $\theta$.

Integrating over the solid angle yields the effective solid angle as a function of core distance (upper right plot):
\begin{equation}
\Omega(r) = \int_0^{2\pi} \int_0^{\pi/2} \varepsilon(\theta, r) \sin\theta d\theta d\phi
\end{equation}
The effective solid angle can accurately be parameterized using the exponentially modified Gaussian distribution:
\begin{align}
\label{eq:omega}
\Omega(r; \alpha, \mu, \sigma, \lambda) = & \quad \alpha \exp \left[\frac{\lambda}{2}(2\mu + \lambda \sigma^2 - 2r)\right] \nonumber\\
&\times \mathrm{erfc}\left( \frac{\mu + \lambda \sigma^2 -r}{\sqrt{2}\sigma} \right)
\end{align}
with $\alpha$ a scaling parameter, $\mu$ and $\sigma$ the mean and standard deviation of the Gaussian part of the distribution and $\lambda$ the rate of the exponential part. The complementary error function, $\mathrm{erfc}(x)$, is given by:
\begin{equation}
\textup{erfc}(x) = \frac{2}{\sqrt{\pi}}\int_x^\infty e^{-y^2}dy
\end{equation}
Integrating the detection efficiency over the surface area (polar coordinates $r$ and $\zeta$) yields the effective surface area as a function of zenith angle (bottom right plot in figure \ref{fig:solid_angle_surface_area}):
\begin{equation}
\textup{A}(\theta) = \int_0^{2\pi} \int_0^{\infty} \varepsilon(\theta, r) r dr d\zeta
\end{equation}
The effective surface area can be parameterized using the following formula:
\begin{equation}
\label{eq:a}
\textup{A}(\theta) = a \exp \left( -b \left( \frac{1}{\cos \theta} - 1 \right) \right)
\end{equation}
with $a$ and $b$ fit parameters. Since, in order to obtain the flux, the number of events needs to be divided by both the effective solid angle and the effective surface area; there is no need to evaluate them separately. Instead, the two are combined:
\begin{align}
\label{eq:omegaint}
\textup{A}\Omega &= \int_0^{2\pi} \int_0^{\infty} \Omega(r) r dr d\zeta \\
\label{eq:aint}
&= \int_0^{2\pi} \int_0^{\pi/2} A(\theta) \sin\theta d\theta d\phi
\end{align}
Table \ref{tab:fitvalues} shows the fit parameters that describe the effective solid angle and surface area for the simulated energies (for $10^{15}$ eV proton showers $\textup{A}\Omega$ is $8.58 \cdot 10^{3}$ \si{\meter\squared\steradian}). The $\textup{A}\Omega$ values can be calculated from eqs. \ref{eq:omegaint} or \ref{eq:aint}, or by direct summation of the two-dimensional histogram in figure \ref{fig:solid_angle_surface_area}. The last column in table \ref{tab:fitvalues} lists the averages and standard deviations of these three methods.

\begin{figure}
	\centering
	\includegraphics[width=90 mm]{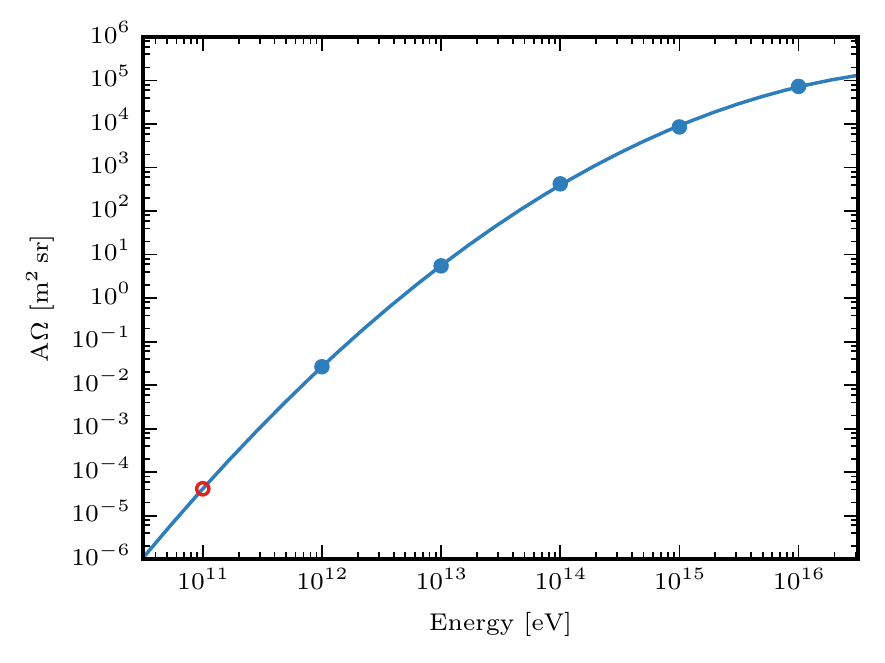}
	\caption{The blue curve (eq. \ref{eq:aomegas}) shows the interpolation of the $\textup{A}\Omega$ values (blue dots) listed in in table \ref{tab:fitvalues}. $\textup{A}\Omega$ at $10^{11}$ eV (red circle, $\textup{A}\Omega=4.2\cdot 10^{-5}$ \si{\meter\squared\steradian}) is obtained by extrapolation.}
	\label{fig:aomega_interpolation}
\end{figure}

Figure \ref{fig:aomega_interpolation} shows the five $\textup{A}\Omega$ values. The interpolation is defined by the following equation:
\begin{equation}
\label{eq:aomegas}
\log_{10}(\textup{A}\Omega) = a x^2 + b x + c
\end{equation}
with $x = \log_{10}(E)$, $a=-0.239$, $b=8.31$, $c=-66.8$. Extrapolation leads to an estimate at $10^{11}$ eV of $\textup{A}\Omega=4.2\cdot 10^{-5}$ (red circle). A precise direct estimate at this energy from simulation is difficult because it is computationally expensive to collect a sufficiently large data set; the number of EASs that satisfy the selection criteria is extremely small. The number of triggers at an EAS energy of $10^{11}$ eV is approximately 400 times smaller than at $10^{12}$ eV. This is consistent with the estimate obtained by extrapolation of eq. \ref{eq:aomegas}.

\begin{table*}
\centering
\caption{The parameters in eq. \ref{eq:omega} (effective solid angle $\Omega(r)$) and eq. \ref{eq:a} (effective surface area $\textup{A}(\theta)$) are listed as a function of energy. The last column gives the value of the combination $\textup{A}\Omega$ (eq. \ref{eq:omegaint} or \ref{eq:aint}).}
\label{tab:fitvalues}
\begin{tabular}{l l l l l l l l}
\hline
Energy [eV] & $\Omega(r)$ & & & & $\textup{A}(\theta)$ & & $\textup{A}\Omega$ [\si{\meter\squared\steradian}] \\
& $\alpha$ & $\mu$ & $\sigma$ & $\lambda$ & $a$ & $b$ & \\
\hline
$10^{12}$ & $1.73 \cdot 10^{-5}$ & $-2.3$ & 1.0 & $8.00 \cdot 10^{-2}$ & $4.86 \cdot 10^{-2}$ & 9.66 & $(2.75 \pm 0.09) \cdot 10^{-2}$ \\
$10^{13}$ & $8.71 \cdot 10^{-3}$ & $-3.1$ & 3.6 & $1.22 \cdot 10^{-1}$ & 7.85& 6.89 & $5.50 \pm 0.14$ \\
$10^{14}$ & $5.44 \cdot 10^{-1}$ & $-10.4$ & 8.1 & $8.78 \cdot 10^{-2}$ & $6.30 \cdot 10^{2}$ & 7.90 & $(4.23 \pm 0.21) \cdot 10^{2}$ \\
$10^{15}$ & 4.40 & $-14.2$ & 24.5 & $7.25 \cdot 10^{-2}$ & $9.21 \cdot 10^{3}$ & 4.76 & $(8.58 \pm 0.28) \cdot 10^{3}$ \\
$10^{16}$ & 6.92 & $-50.7$ & 63.3 & $2.71 \cdot 10^{-2}$ & $6.33 \cdot 10^{4}$ & 3.62 & $(7.33 \pm 0.13) \cdot 10^{4}$ \\
\hline
\end{tabular}
\end{table*}

\section{Fitting single energy PIDs to experimental data}
The data set used for figure \ref{fig:timediff} was used to obtain a PID. The contribution to the PID from random coincidences was subtracted (see section \ref{sec:detsep}). A Bayesian method was used to fit a combination of simulated single energy PIDs in the range $10^{12}$$-$$10^{16}$ eV to the experimental data. Only pulse integrals smaller than 35 Vns are considered to limit systematic differences between the detectors (figure \ref{fig:systematics}). Signals are affected by gamma rays, and Cherenkov light generated in the light-guide \cite{vandam2020}. Especially at lower multiplicities this contribution becomes apparent. Pulse integrals below 6 Vns (2 MIPs) are therefore discarded as well.

As shown in figure \ref{fig:aomega_interpolation}, the probability that a $10^{11}$ eV shower results in a footprint that triggers the station is negligible. The slope of PIDs with energies $10^{17}$ eV and beyond becomes rather similar to that of $10^{16}$ eV in the pulse integral range between 6 to 35 Vns whereas the flux at those energies rapidly decreases. The experimentally observed PID will therefore be restricted to the sum of five single energy contributions ($10^{12}$$-$$10^{16}$ eV):
\begin{align}
z(n_i,\vec{x}) = &\ x_1 \cdot \textup{PID}_{12}(n_i) + x_2 \cdot \textup{PID}_{13}(n_i) \nonumber \\
&+ x_3 \cdot \textup{PID}_{14}(n_i) + x_4 \cdot \textup{PID}_{15}(n_i) \nonumber \\
&+ x_5 \cdot \textup{PID}_{16}(n_i)
\end{align}
Here $\vec{x}\equiv(x_1, x_2, x_3, x_4, x_5)$, $n_i$ is the bin value, $z(n_i)$ is the expected number of events in each bin and each $\textup{PID}(n_i)$ is a single energy model (fig. \ref{fig:ph_per_energy}). The uncertainty in the number of entries in the bins of the experimentally observed PID is described by a Poisson distribution. The probability to obtain a number of counts $y_i$ in bin $n_i$ given $\vec{x}$ is given by:
\begin{equation}
p(y_i,n_i|\vec{x})=P(y_i,z(n_i,\vec{x}))
\end{equation}
with the Poisson distribution:
\begin{equation}
P(y,z)=\frac{z^y}{y!}e^{-z}
\end{equation}

Bayes' rule can be exploited to define a probability to obtain the fit parameters $\vec{x}$.
\begin{equation}
p(\vec{x}|N,Y)=\frac{p(N,Y|\vec{x})p(\vec{x})}{p(N,Y)}
\end{equation}
Here $N$ and $Y$ are the combined collection of elements $n_i$ and $y_i$. The denominator $p(N,Y)$ is a normalization constant. The $p(\vec{x})$ is known as the prior probability distribution. The prior states that the parameters $x_1$, $x_2$, $x_3$, $x_4$ and $x_5$ cannot become negative. The function $p(\vec{x}|N,Y)$ is the posterior probability distribution. The $p(N,Y|\vec{x})$ function is the likelihood (the product of conditional probabilities):
\begin{equation}
p(N,Y|\vec{x})=\prod_{i=1}^M p(y_i,n_i|\vec{x}) = \mathcal{L}
\end{equation}

Since the slope differences between the PIDs occur at higher multiplicities, a gradually larger weight is assigned with increasing multiplicity. This leads to a weighted likelihood:
\begin{equation}
\mathcal{\widehat{L}}=\prod_{i=1}^M p(y_i,n_i|\vec{x})^{w(n_i)}
\end{equation}
with $w(n_i)$ the weight function:
\begin{equation}
w(n_i)=1.161^{n_i}
\end{equation}
which corrects for the number of entries in the lower ($1.7 \cdot 10^4$ at 6 Vns) and higher bins (231 at 35 Vns). By taking the logarithm of this weighted likelihood, the product converts into:
\begin{align}
\log(\mathcal{\widehat{L}}) &=  \sum_{i=1}^M w(n_i) \cdot \log \left( \frac{{z_i}^{y_i}}{y_i!}e^{-z_i} \right)\\
&= \sum_{i=1}^M w(n_i) \cdot (y_i \log(z_i) - z_i - \log(y_i!)) \\
&= \sum_{i=1}^M w(n_i) \cdot (y_i \log(z_i) - z_i - C) 
\end{align}
with $z_i=z(n_i,\vec{x})$ and $C$ is a constant.

\begin{figure}
	\centering
	\includegraphics[width=90 mm]{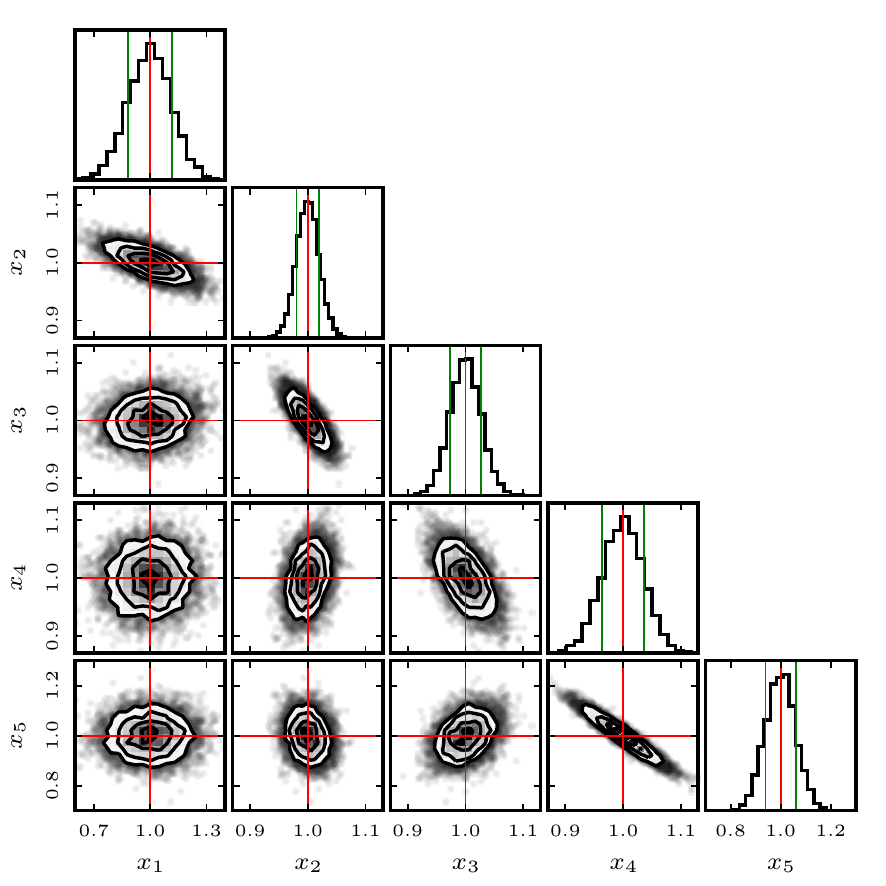}
	\caption{Subsets of the sampled posterior probability distributions. The upper Gaussian shaped histograms show the sample selection in the dimension of the fit parameters ($\vec{x}$). The median of the histogram (red lines) is taken as the best fit. The uncertainty in the fit is shown by the standard deviation (green lines). The other subplots display the relation between two fit parameters. There is some interdependence between the fit parameters of neighboring energies. This is more pronounced for $x_4$ and $x_5$ (fourth sub-plot in bottom row). See also figure \ref{fig:ph_per_energy}.}
	\label{fig:corner}
\end{figure}

\begin{figure}
	\centering
	\includegraphics[width=90 mm]{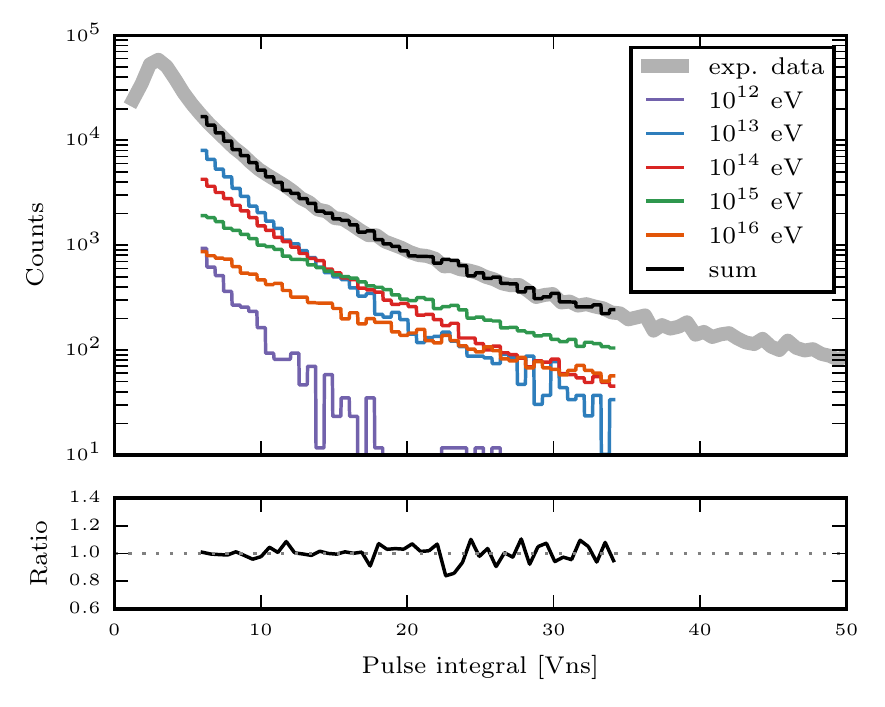}
	\caption{A linear combination of five simulated PIDs (colored lines, see figure \ref{fig:ph_per_energy}) fitted to the experimentally obtained PID (thick grey line). The black line shows the best fit. The bottom plot gives the ratio of the experimental data and the fit.}
	\label{fig:fit_procedure}
\end{figure}

Instead of directly maximizing the posterior probability distribution, a range of parameters ($\vec{x}$) are explored. This is done using a Markov Chain Monte Carlo (MCMC) algorithm \cite{foreman2013}. Figure \ref{fig:corner} shows (lower dimensional) subsets of the sampled (five dimensional) posterior probability distribution. The upper (on the diagonal) Gaussian shaped histograms show the sample selection in the dimension of the fit parameters. The median of the histogram (red lines) is taken as the best fit. The uncertainty is shown by the standard deviation (green lines). The best fit values are $(3.01 \pm 0.35) \cdot 10^4$, $(2.32 \pm 0.05) \cdot 10^5$, $(1.09 \pm 0.03) \cdot 10^5$, $(5.63 \pm 0.20) \cdot 10^4$, $(2.63 \pm 0.16) \cdot 10^4$ for $10^{12}$ to $10^{16}$ eV resp. The other subplots display the relation between two fit parameters. There is some interdependence between the fit parameters of neighboring energies. This is especially pronounced at fit parameters $x_4$ and $x_5$ (fourth sub-plot in bottom row) due to the relatively small slope differences between the $10^{15}$ and $10^{16}$ eV PIDs (figure \ref{fig:ph_per_energy}). The resulting fit to the experimental data is shown in figure \ref{fig:fit_procedure}.

In addition to the statistical uncertainty in the experimental data, the single energy models also have an intrinsic statistical uncertainty. Creating larger data sets for the single energy PIDs is currently limited by the generation of simulated EASs which is computationally expensive. The uncertainty has been estimated by resampling the single energy PIDs. The number of events in each bin was randomly resampled following Poisson statistics. A combination of these new PIDs was fitted to the experimental data as well. This procedure was carried out multiple times. The mean and standard deviation of the best fit parameters are listed in table \ref{tab:numberofevents}.

\begin{table}
\centering
\caption{Best fit for the number of events with their uncertainties.}
\label{tab:numberofevents}
\begin{tabular}{l l}
\hline
Energy [eV] & Number of events \\
\hline
$10^{12}$ & $(4.09 \pm 1.76) \cdot 10^4$ \\
$10^{13}$ & $(2.04 \pm 0.23) \cdot 10^5$ \\
$10^{14}$ & $(1.25 \pm 0.20) \cdot 10^5$ \\
$10^{15}$ & $(5.65 \pm 1.32) \cdot 10^4$ \\
$10^{16}$ & $(2.43 \pm 1.04) \cdot 10^4$ \\
\hline
\end{tabular}
\end{table}

\section{Cosmic ray flux}
\begin{figure}
\centering
\includegraphics[width=90 mm]{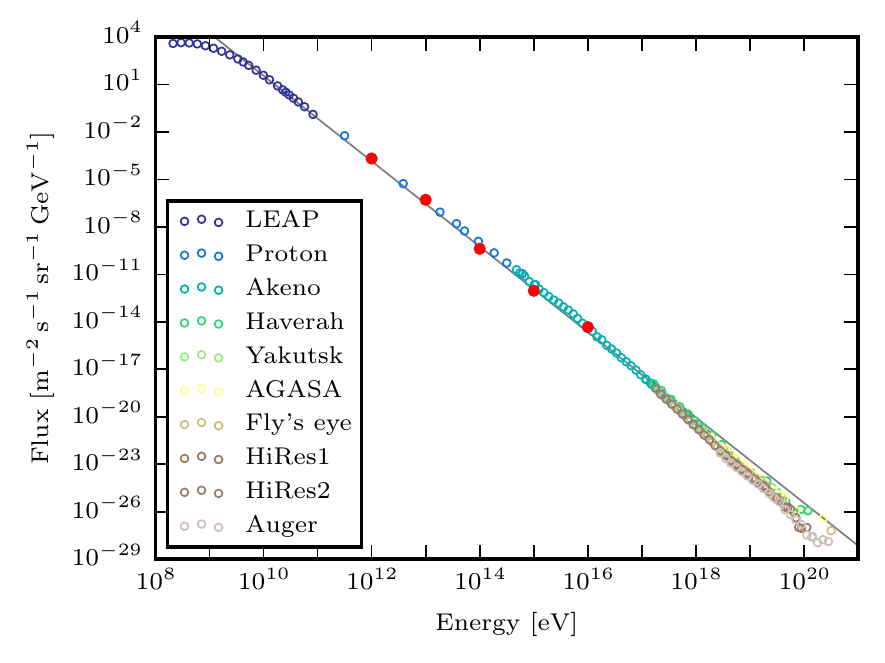}
\caption{Cosmic ray fluxes at $10^{12}$$-$$10^{16}$ eV from HiSPARC data (red dots) compared to results from other experiments \cite[and references therein]{spectrum, cronin1997} (circles). The uncertainties are smaller than the dot size. The fluxes obtained by HiSPARC agree well with the other measurements. The grey line represents eq. \ref{eq:crflux}.}
\label{fig:energy_spectrum}
\end{figure}

The estimated number of events per energy decade derived from the fit can be used in combination with the effective surface area and solid angle to obtain the cosmic ray flux as a function of energy. The fluxes are calculated using:
\begin{equation}
\label{eq:fluxcalc}
F = \frac{N_\textup{events}}{\textup{A}\Omega \cdot t \cdot \Delta E}
\end{equation}
with $t$ the duration of the experiment ($t=29$ days $=2.506 \cdot 10^6$ s) and $\Delta E$ the width of the energy bin (e.g. $10^{15.5} - 10^{14.5} \approx 2.85 \cdot 10^{15}$ eV). The flux values are listed in table \ref{tab:fluxvalues}. Figure \ref{fig:energy_spectrum} shows the cosmic ray energy spectrum (circles) from several experiments \cite[and references therein]{spectrum, cronin1997} together with the flux values derived from HiSPARC data (red dots). The uncertainties are smaller than the dot size. The grey line represents the function in equation \ref{eq:crflux}. The HiSPARC values agree well with the other measurements. Note that the reference flux values below $10^{14}$ eV are measured using spacecraft (e.g. Proton satellite \cite{grigorov1971}).

\begin{table}
\centering
\caption{Cosmic ray flux values obtained using a two-detector HiSPARC station.}
\label{tab:fluxvalues}
\begin{tabular}{l l}
\hline
Energy [eV] & Flux [\si{\per\meter\squared\per\second\per\steradian\per\giga\electronvolt}] \\
\hline
$10^{12}$ & $(2.08 \pm 0.93) \cdot 10^{-4}$ \\
$10^{13}$ & $(5.19 \pm 0.60) \cdot 10^{-7}$ \\
$10^{14}$ & $(4.14 \pm 0.70) \cdot 10^{-10}$ \\
$10^{15}$ & $(9.22 \pm 2.20) \cdot 10^{-13}$ \\
$10^{16}$ & $(4.64 \pm 2.01) \cdot 10^{-15}$ \\
\hline
\end{tabular}
\end{table}

\section{Discussion and conclusion}
A HiSPARC station (at sea-level) with only two 0.5 \si{\meter\squared} scintillator detectors was used to probe the cosmic ray energy spectrum in the range $10^{12}$$-$$10^{16}$ eV (five energy decades). One month of data shows to be sufficient to obtain flux values that are in good agreement with the results from dedicated (space-based) experiments.

The presented analysis is very different from those applied by other ground-based cosmic ray observatories. The method does not require the reconstruction of the shower core and/or shower size of individual EASs but relies on deriving the energy dependent particle multiplicity distributions in a single scintillator detector. PMT pulse integrals are used as a proxy to estimate the number of particles simultaneously traversing the detector. In comparing simulated PIDs at $10^{15}$ and $10^{16}$ eV (and beyond $10^{16}$ eV) with experimental data, the simulation starts to suffer from lack of statistics; showers become very large exceeding the presently available CPU power and data storage facilities. Re-doing the analysis with simulations in which 'thinning' is applied has not been explored but may offer an alternative. On the other hand, as becomes apparent from figure \ref{fig:ph_per_energy}, the difference in slope of the PIDs towards higher particle multiplicities decreases with increasing energy. Moreover, at these large multiplicities, the dynamical range of the readout electronics and non-linearities in detector response require careful evaluation. By extending the distance between the detectors, the contribution from small, low energy ($10^{12}$$-$$10^{13}$ eV) EASs can be reduced, improving the sensitivity of the analysis towards higher energy EASs.

Since the HiSPARC network consists of more than 120 stations in which the separation between the scintillators varies between 5 and 17 m while for several stations more than 10 years of data are stored, statistics can easily be increased and systematic uncertainties can be investigated in detail.

Finally, the cosmic ray flux below $10^{12}$ eV could possibly be derived by increasing the altitude of the HiSPARC station, compensating for the shift of the shower maximum towards higher altitudes. The two-detector station in Windhoek (Namibia) is at an altitude of almost 2 km...

\section*{Acknowledgements}
We thank Hogeschool lecturer and high school teacher Kees van der Velden for carrying out the measurements at the roof of his high school presented in figure \ref{fig:pi_distances}. Also the support of the contributing HiSPARC schools, their teachers and especially their students is gratefully acknowledged. Without their help, it would have been impossible to keep the stations 'up and running'. 



\bibliographystyle{elsarticle-num}
\bibliography{references.bib}

\begin{thebibliography}{10}
\expandafter\ifx\csname url\endcsname\relax
  \def\url#1{\texttt{#1}}\fi
\expandafter\ifx\csname urlprefix\endcsname\relax\def\urlprefix{URL }\fi
\expandafter\ifx\csname href\endcsname\relax
  \def\href#1#2{#2} \def\path#1{#1}\fi

\bibitem{gaisser2016}
T.~K. Gaisser, R.~Engel, E.~Resconi, {Cosmic rays and particle physics},
  Cambridge University Press, 2016.

\bibitem{physicalreviewd2012}
J.~Beringer, J.~Arguin, R.~Barnett, K.~Copic, O.~Dahl, D.~Groom, C.~Lin,
  J.~Lys, H.~Murayama, C.~Wohl, et~al., Review of particle physics, Physical
  Review D-Particles, Fields, Gravitation and Cosmology 86~(1) (2012) 010001.

\bibitem{corsika1998}
D.~Heck, J.~Knapp, J.~N. Capdevielle, et~al., {CORSIKA: a Monte Carlo code to
  simulate extensive air showers.}, Forschungszentrum Karlsruhe GmbH, Karlsruhe
  (Germany), 1998.

\bibitem{vandam2020}
K.~van Dam, B.~van Eijk, D.~Fokkema, et~al., The {HiSPARC} experiment, Nuclear
  Instruments and Methods in Physics Research Section A: Accelerators,
  Spectrometers, Detectors and Associated Equipment 959 (2020) 163577.

\bibitem{GEANT42016}
J.~Allison, K.~Amako, J.~Apostolakis, et~al., {Recent developments in Geant4},
  Nucl. Instrum. Methods Phys. Res. A 835 (2016) 186--225.

\bibitem{vandam2020b}
K.~van Dam, Increased radioactivity during precipitation measured by the
  {HiSPARC} experiment, Physica Scripta 95 (2020) 074011.

\bibitem{landau1944}
L.~Landau, {On the energy loss of fast particles by ionization}, J. Phys.(USSR)
  8 (1944) 201--205.

\bibitem{takeda2003}
M.~Takeda, N.~Sakaki, K.~Honda, M.~Chikawa, M.~Fukushima, N.~Hayashida,
  N.~Inoue, K.~Kadota, F.~Kakimoto, K.~Kamata, et~al., Energy determination in
  the akeno giant air shower array experiment, Astroparticle Physics 19 (2003)
  447--462.

\bibitem{dai1988}
H.~Dai, K.~Kasahara, Y.~Matsubara, M.~Nagano, M.~Teshima, On the energy
  estimation of ultra-high-energy cosmic rays observed with the surface
  detector array, Journal of Physics G: Nuclear Physics 14 (1988) 793.

\bibitem{qgsjet2006}
S.~Ostapchenko, {QGSJET-II: towards reliable description of very high energy
  hadronic interactions}, in: Nuclear Physics B - Proceedings Supplements, Vol.
  151, 2006, pp. 143--146.

\bibitem{gheisha1985}
H.~Fesefeldt, {GHEISHA The Simulation of Hadronic Showers}, Tech. rep., RWTH
  Aachen, {PITHA-85/02} (1985).

\bibitem{egs41985}
W.~R. Nelson, D.~W.~O. Rogers, H.~Hirayama, {The EGS4 code system}, Tech. rep.,
  Stanford Linear Accelerator Center, Stanford, California (1985).

\bibitem{hillas1981}
A.~M. Hillas, {}, in: Proceedings International Cosmic Ray Conference, Vol.~1,
  1981, p. 193.

\bibitem{foreman2013}
D.~Foreman-Mackey, D.~W. Hogg, D.~Lang, J.~Goodman, emcee: the mcmc hammer,
  Publications of the Astronomical Society of the Pacific 125 (2013) 306.

\bibitem{spectrum}
{Cosmic Ray Spectrum website by Dr. William Hanlon},
  \url{https://www.physics.utah.edu/~whanlon/spectrum.html}.

\bibitem{cronin1997}
J.~W. Cronin, T.~K. Gaisser, S.~P. Swordy, Cosmic rays at the energy frontier,
  Scientific American 276 (1997) 44--49.

\bibitem{grigorov1971}
N.~Grigorov, Y.~V. Gubin, I.~Rapoport, I.~Savenko, B.~Yakovlev, V.~Akimov,
  V.~Nesterov, Energy spectrum of primary cosmic rays in the 10 11-10 15 ev
  energy range according to the data of proton-4 measurements, in: 12th
  International conference on cosmic rays, 16-25th August 1971, Hobart,
  Tasmania, Australia, 1971.

\end{thebibliography}

\end{document}